\documentclass[conference]{IEEEtran}
\IEEEoverridecommandlockouts
\usepackage{cite}
\usepackage{balance}
\usepackage{amsmath,amssymb,amsfonts}
\usepackage{algorithmic}
\usepackage{graphicx}
\usepackage{textcomp}
\usepackage{xcolor}
\usepackage{array}
\usepackage{pifont}
\usepackage{enumitem}
\usepackage{graphicx}
\usepackage{url}
\usepackage{multirow}
\usepackage{pifont}
\usepackage{tikz}
\newcommand*\circled[1]{\tikz[baseline=(char.base)]{\node[shape=rectangle,draw,inner sep=2pt] (char) {#1};}}

\def\BibTeX{{\rm B\kern-.05em{\sc i\kern-.025em b}\kern-.08em
    T\kern-.1667em\lower.7ex\hbox{E}\kern-.125emX}}
\begin{document}

\title{SE-Enhanced ViT and BiLSTM-Based Intrusion Detection for Secure IIoT and IoMT Environments}

%Enhanced Vision-Transformer and BiLSTM-based IDS for Securing IoMT and IIoT networks

\author{\IEEEauthorblockN{1\textsuperscript{st} Afrah Gueriani}
\IEEEauthorblockA{\textit{LSEA Lab., Faculty of Technology} \\
\textit{University of MEDEA}\\
Medea 26000, Algeria \\
gueriani.afrah@univ-medea.dz }
\and
\IEEEauthorblockN{2\textsuperscript{nd} Hamza Kheddar}
\IEEEauthorblockA{\textit{LSEA Lab., Faculty of Technology} \\
\textit{University of MEDEA}\\
Medea 26000, Algeria\\
kheddar.hamza@univ-medea.dz}
\and
\IEEEauthorblockN{3\textsuperscript{rd} Ahmed Cherif Mazari}
\IEEEauthorblockA{\textit{LSEA Lab, Faculty of Science} \\
\textit{University of MEDEA}\\
Medea 26000, Algeria \\
mazari.ahmedcherif@univ-medea.dz}
\and
\IEEEauthorblockN{4\textsuperscript{th} Seref Sagiroglu}
\IEEEauthorblockA{\textit{Department of Computer Engineering} \\
\textit{Gazi University}\\
Ankara, Turkey \\
ss@gazi.edu.tr}
\and
\IEEEauthorblockN{5\textsuperscript{th} Onur Ceran}
\IEEEauthorblockA{\textit{Department of Management Information Systems} \\
\textit{Gazi University}\\
Ankara, Turkey \\
onur.ceran@gazi.edu.tr}
}

\maketitle

\begin{abstract}
With the rapid growth of interconnected devices in Industrial and Medical Internet of Things (IIoT and MIoT) ecosystems, ensuring timely and accurate detection of cyber threats has become a critical challenge. This study presents an advanced intrusion detection framework based on a hybrid Squeeze-and-Excitation Attention Vision Transformer-Bidirectional Long Short-Term Memory (SE ViT-BiLSTM) architecture. In this design, the traditional multi-head attention mechanism of the Vision Transformer is replaced with Squeeze-and-Excitation attention, and integrated with BiLSTM layers to enhance detection accuracy and computational efficiency. 
The proposed model was trained and evaluated on two real-world benchmark datasets; EdgeIIoT and CICIoMT2024; both before and after data balancing using the Synthetic Minority Over-sampling Technique (SMOTE) and RandomOverSampler. Experimental results demonstrate that the SE ViT-BiLSTM model outperforms existing approaches across multiple metrics. Before balancing, the model achieved accuracies of 99.11\% (FPR: 0.0013\%, latency: 0.00032 sec/inst) on EdgeIIoT and 96.10\% (FPR: 0.0036\%, latency: 0.00053 sec/inst) on CICIoMT2024. After balancing, performance further improved, reaching 99.33\% accuracy with 0.00035 sec/inst latency on EdgeIIoT and 98.16\% accuracy with 0.00014 sec/inst latency on CICIoMT2024.

\end{abstract}

\begin{IEEEkeywords}
Squeeze-and-Excitation attention, Vision Transformer, Bidirectional Long Short-Term Memory, Intrusion Detection System, Industrial Internet of Things, Internet of Medical Things, Cybersecurity, Network traffic analysis.

\end{IEEEkeywords}

\section{Introduction}
The rapid expansion of the Internet of Things (IoT) has transformed various industries by enabling seamless connectivity, efficient data collection, and automation \cite{kamir2023machine}. This has led to the development of specialized domains such as the Industrial Internet of Things (IIoT) and the Internet of Medical Things (IoMT), which enhance operational efficiency through automated data analysis, predictive maintenance, and informed decision-making \cite{ni2024machine,gueriani2025cyber,gueriani2025explainable}. IoMT systems, in particular, enable real-time patient data collection and transmission to healthcare professionals, improving diagnostic accuracy and reducing human error \cite{ahmed2024insights, xu2023efficient}. However, the heterogeneous and highly connected nature of these networks increases vulnerability to cyber threats, putting data confidentiality, availability, and system integrity at risk \cite{wang2023securing}. As IoT becomes more integrated into critical infrastructure, the need for robust and intelligent Intrusion Detection Systems (IDS) becomes essential to maintain security and reliability in dynamic IoMT and IIoT environments \cite{sana2024securing}. Traditional security mechanisms often struggle to detect advanced or evolving cyber threats. Intrusion Detection Systems (IDS) address this challenge by continuously monitoring network activity and analyzing abnormal behavior \cite{buyuktanir2025federated}. Common IDS techniques in IoT include signature-based, anomaly-based, and behavior-based detection. Anomaly detection, in particular, plays a key role in identifying potential threats by analyzing network traffic for unusual patterns \cite{gandi2023comparative}, enabling timely security responses \cite{gandi2023comparative, ali2022applied}. To enhance IDS capabilities, this study proposes a hybrid deep learning (DL) model that combines a squeeze-and-excitation (SE) enhanced Vision Transformer (ViT) with a Bidirectional Long Short-Term Memory (BiLSTM) network. This architecture leverages ViT’s global feature modeling, BiLSTM’s temporal learning, and SE’s adaptive feature recalibration to improve detection accuracy. 

\subsection{Our Contribution}

The key contributions of our work include:

\begin{itemize}
\item Design a novel hybrid IDS architecture that integrates SE blocks with a ViT and BiLSTM, aiming to enhance both spatial feature extraction and temporal dependency modeling.
\item Evaluate the proposed model using two benchmark datasets: the Edge-IIoT2024 dataset and a subset of the augmented CIC-IoMT2024 dataset.
\item To address class imbalance, SMOTE is applied to the Edge-IIoT2024 dataset, and RandomOverSampler is used for the CIC-IoMT2024 subset. Comparative experiments are performed before and after balancing.
\item The study targets a multiclass classification task, aiming to accurately detect and classify various cyber attack types across different IoT environments.
\item An in-depth evaluation is conducted using key performance metrics, showing the model’s robustness and effectiveness in identifying complex attack patterns compared to state-of-the-art methods.
\end{itemize}

\subsection{Organization of the paper}
The rest of the paper is structured as follows: Section \ref{sec2} presents the problem statement and similar works from the literature. Section \ref{sec3} provides the proposed methodology and dataset prepossessing. Section \ref{sec4} discusses the experimentation, results, and discussion. Section \ref{sec5} concludes the paper and mentions future work.

\section{Problem Statement and Literature Review}
\label{sec2}
The rapid expansion of interconnected IoT devices has created major security challenges due to their diverse and resource-limited nature. These systems are highly susceptible to threats such as malware, unauthorized access, and data breaches. The core issues include the dynamic and heterogeneous structure of IoT environments, constantly changing network topologies, and the need to both prevent attacks and minimize damage from successful intrusions.

In this context, several state-of-the-art works, have addressed these issues by developing advanced solutions and methodologies, such as the study by 
 O. Ceran et al. \cite{ceran2025leveraging}, propose an innovative hybrid intrusion detection framework that integrates Graph Neural Networks (GNNs) with the XGBoost algorithm to effectively address the evolving security challenges in IoT environments. The experimental evaluation conducted on four real-world datasets namely: CICIoT-2023, CICIDS-2017, UNSW-NB15, and IoMT-2024,  demonstrates that the proposed hybrid model consistently outperforms conventional machine learning (ML) approaches across various performance metrics. 
 Considering \cite{peng2024bilstm}, H. Peng et al. propose a BiLSTM-based IDS tailored for IoT environments. To enhance performance, the authors integrate mutual information for feature selection and apply focal loss to address class imbalance during training.  
 Evaluated on EdgeIIoT benchmark dataset, the proposed IDS demonstrates superior accuracy and robustness compared to traditional methods. Another work by 
A. Gueriani et al. in \cite{gueriani2025robust} propose a robust cross-domain IDS that integrates BiGRU, LSTM, and attention mechanisms to enhance security in IoMT and IIoT environments, and in \cite{gueriani2025hybrid}, the same authors employed ResNet-1D-BiGRU with multi-head Attention. 
Evaluated on Edge-IIoTset and CICIoMT2024 datasets before and after  balancing techniques, the architecture achieves high accuracy across domains, demonstrating strong generalization and adaptability for both schemes. 
However, \cite{gueriani2025robust} highlights its potential for zero-day attacks in diverse IoT infrastructures and extend the study to explainable artificial intelligence (XAI).

\section{Materials and Method}
\label{sec3}

\subsection{Data preprocessing} 
To ensure consistency and optimize the performance of the proposed intrusion detection models, a standardized data preprocessing pipeline was applied to both the Edge-IIoTset and CICIoMT2024 datasets. The following steps were conducted: \\
\noindent \textbf{\textit{- Label Encoding:}} To support multi-class classification tasks, the \textit{Attack\_type} column was mapped to integer values using a predefined dictionary. \\
\noindent \textbf{\textit{- Categorical Feature Encoding:}} Several protocol-related categorical features were identified and transformed using \textit{LabelEncoder} from the scikit-learn library.\\
\noindent \textbf{\textit{- Feature Selection and Cleaning:}} Non-contributory or redundant columns such as \textit{Attack\_type}, \textit{Attack\_label}, and \textit{frame.time} were excluded from the feature set "During the feature selection phase, the Attack type column was removed from the input features, as it
represents the target class. After feature selection, the Attack type column was reshaped and encoded
separately to serve as the label for model training in the supervised learning process3.\\ 
\noindent \textbf{\textit{-Target Variable Encoding:}} The \textit{Attack\_type} column, used as the classification target was reshaped and label-encoded into a numerical format, preparing it for multi-class classification scenarios. \\
\noindent \textbf{\textit{- Addressing Class Imbalance:}} To mitigate the effects of class imbalance, prevalent in both EdgeIIoT and CICIoMT2024 datasets, the \textit{SMOTE} and \textit{RandomOverSampler} were applied respectively for both datasets.\\
\noindent \textbf{\textit{- Train-Validation Splitting:}} The preprocessed selected sets of the datasets are divided into two segments: 80\% of the preprocessed data was allocated to training our proposed model and 20\% of the preprocessed data was reserved to validate the performance of the enhanced SE-based ViT-BiLSTM model.

\begin{figure}
\centerline{\includegraphics[scale=0.3]{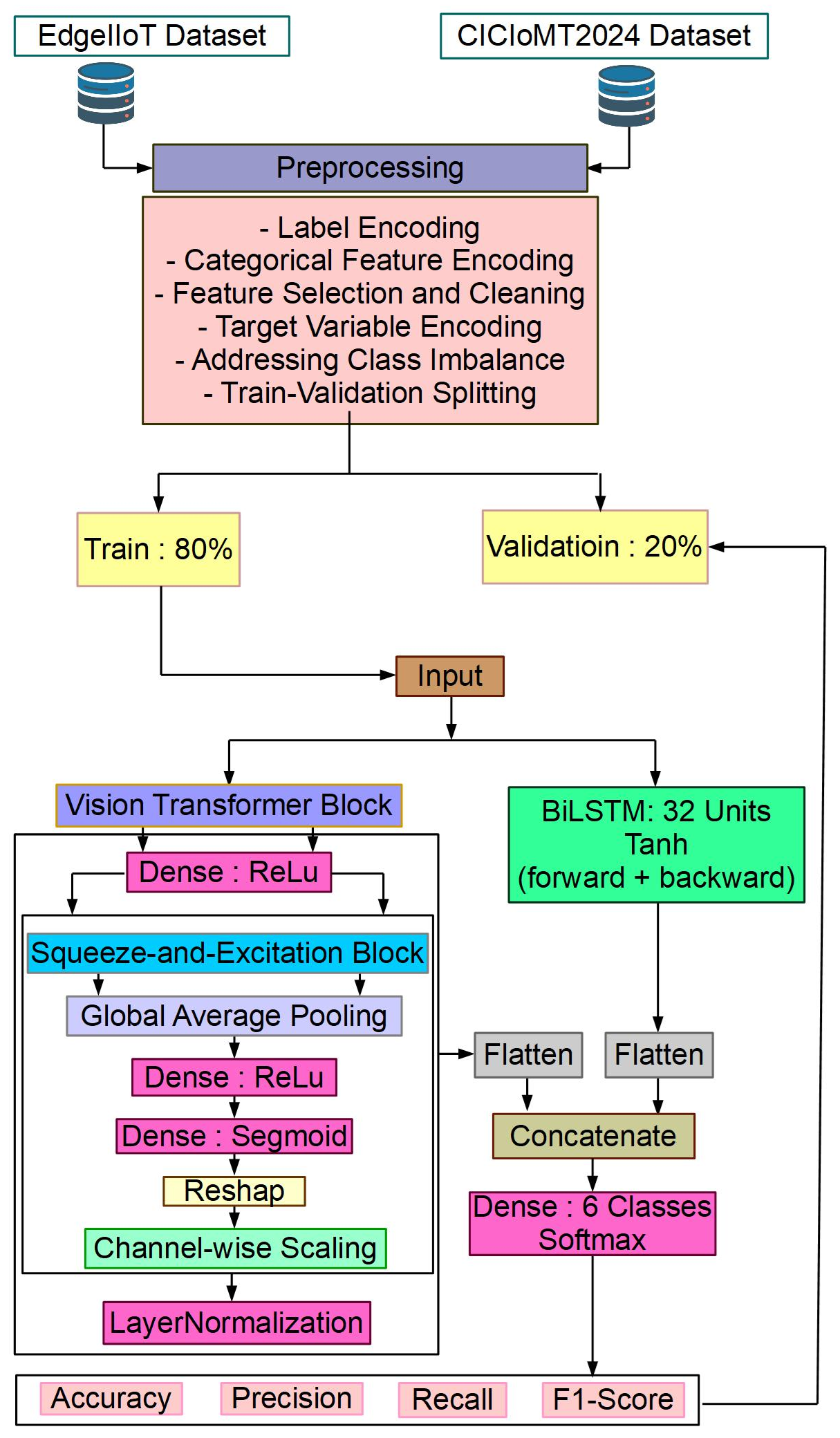}}
\caption{Architectural workflow for Proposed Model.}
\label{fig}
\end{figure}

\subsection{Structural model}

To effectively capture both spatial dependencies and sequential patterns in network traffic data, we propose a hybrid DL architecture that integrates a SE-based ViT block with a BiLSTM network (Figure \ref{fig}).

\begin{itemize}[leftmargin=1em]
    \item \textbf{Input Layer}: The model accepts a dimensional input of shape (60, 1) for EdgeIIoT and (83,1) for CICIoMT2024, where 60 and 83 represent the number of selected features for each dataset and 1 denotes the time dimension.
\item \textbf{Branch 1 (SE-ViT Pathway):} This branch is responsible for feature embedding and attention-based feature recalibration. A \textbf{Dense} layer with ReLU activation projects the input into a higher-dimensional embedding space. A custom \textbf{ViT} block processes the embedded features as a sequence of tokens. However, instead of standard self-attention, we incorporate a \textbf{SE} attention mechanism, which: Performs global average pooling to capture aggregated channel information. than Passes the result through two \textbf{Dense} layers with relu and sigmoid activation to generate channel-wise importance weights. and recalibrates the features by rescaling them with these weights. \\
\noindent - The recalibrated features are added back to the original input through a residual connection and normalized using Layer \textbf{Normalization}. \\
\noindent - The final output is \textbf{flattened} to prepare it for fusion with the BiLSTM pathway.\\
\item \textbf{Branch 2 (BiLSTM Pathway):} This branch captures temporal dependencies in the feature sequence.
\textbf{Bidirectional LSTM} layer with 32 units is used to process the input sequence in both forward and backward directions. The return\_sequences=True setting ensures that the model retains the full temporal dynamics across time steps. The output sequence is then flattened to align with the output from the SE-ViT branch. 
\item \textbf{Feature Fusion and Output Layer:} The \textbf{flattened} outputs from both branches are concatenated to form a comprehensive feature vector that captures both spatial and temporal dependencies. This vector is passed through a \textbf{Dense} output layer with softmax activation to classify the input into one of six output classes.

\end{itemize}

\section{Experimentation, results and discussion}
\label{sec4}

\subsection{Dataset description}
\noindent \textbf{\textit{- EdgeIIoT dataset\footnote{\url{ https://www.kaggle.com/code/mohamedamineferrag/edge-iiotset-pre-processing}}:}} The Edge-IIoTset dataset, introduced in 2022 by Ferrag et al. \cite{ferrag2022edge}, serves as a comprehensive and publicly available benchmark for evaluating cyber security mechanisms in real-world IoT and IIoT environments. 
The dataset features 14 attack types grouped into five major categories.

\noindent \textbf{\textit{- CICIoMT2024 dataset\footnote{\url{https://www.unb.ca/cic/datasets/iomt-dataset-2024.html}}:}} introduced by the Canadian Institute for Cyber security (CIC) in 2024  \cite{dadkhah2024ciciomt2024}, represents a robust and realistic benchmark tailored for the evaluation of IDS within IoMT environments. 
It features 18 distinct attack types, systematically grouped into five major categories.

\begin{figure}[htbp]
\centerline{\includegraphics[scale=0.5]{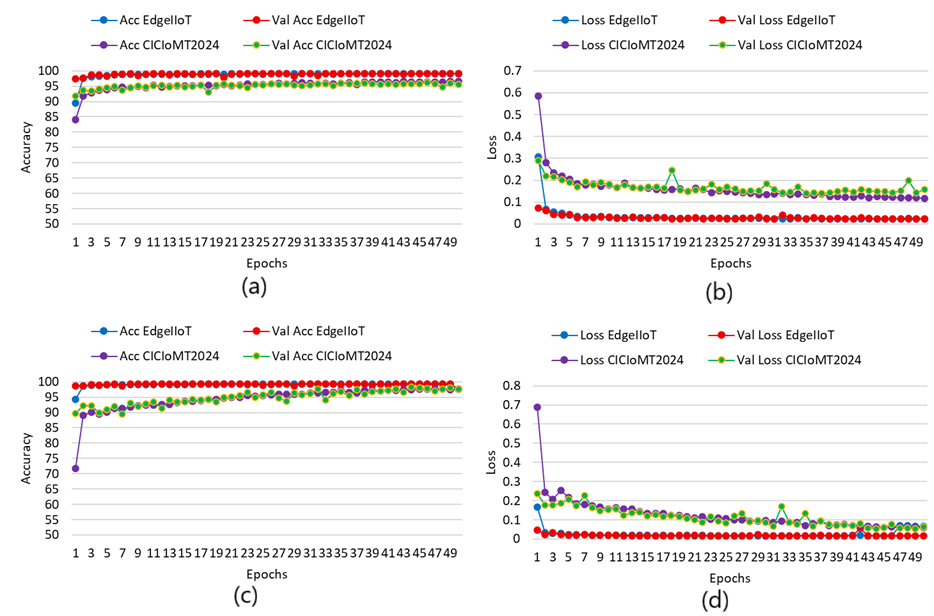}}
\caption{Accuracy and loss of the proposed model. (a): Accuracies before balancing the dataset, (b): Loss before balancing the dataset, (c): Accuracies after balancing the dataset, (d): Loss after balancing the dataset.}
\label{fig2}
\end{figure}

\subsection{Performance metrics}

To comprehensively assess the performance of the proposed SE ViT-BiLSTM model-based IDS, a range of standard classification metrics was employed. These metrics are derived from the fundamental confusion matrix components: True Positives (TP), True Negatives (TN), False Positives (FP), and False Negatives (FN). 
These metrics are formally defined and described in detail in \cite{gueriani2023deep,kheddar2024deep, kheddarASR2023, gueriani2024adaptive, gueriani2024enhancing}.

\begin{table}[ht!]
\centering
\caption{Comprehensive Evaluation Results of the Enhanced SE ViT-BiLSTM Model Using EdgeIIoT and CICIoMT2024 Datasets.}
\scriptsize
\vspace{0.1cm}
\label{tab6}
%\begin{tabular}{|c|c|c|c|c|c|c|c|c|}
\begin{tabular}{|cm{1.35cm}m{0.35cm}m{0.5cm}m{0.35cm}m{0.35cm}m{0.35cm}m{0.5cm}m{0.7cm}|}
\hline
B & Dataset& Acc & Loss & Pr  & Rc & F1  & FPR & Inf time\\
&  &  (\%)  &  & (\%) &(\%) & (\%) &(\%) & (sec/ins)\\
\hline
\hline
No & \textbf{EdgeIIoT} &\textbf{99.11} & \textbf{0.0247} & \textbf{99.11} & \textbf{99.11} & \textbf{99.11} & \textbf{0.0013}& \textbf{0.00032}\\[0.6mm]

& \textbf{CICIoMT2024}&\textbf{96.10} & \textbf{0.1440} & \textbf{96.10} & \textbf{96.10} & \textbf{96.10} & \textbf{0.0223} & \textbf{0.00053}\\

\hline
%\hline

Yes & \textbf{EdgeIIoT} &\textbf{99.33} & \textbf{0.0158} & \textbf{99.33} & \textbf{99.33} & \textbf{99.33} & \textbf{0.0013}& \textbf{0.00035}\\[0.6mm]

& \textbf{CICIoMT2024} &\textbf{98.16} & \textbf{0.0578} & \textbf{98.16} & \textbf{98.16} & \textbf{98.15} & \textbf{0.0036} & \textbf{0.00014}\\
\hline
\end{tabular}
\begin{flushleft}
 Abbreviation: Balancing (B)   
\end{flushleft}
\end{table}

\subsection{Experiments and results} 
The experiments used the Edge-IIoTset and CICIoMT2024 datasets, trained on Google Colab with a T4 GPU and 12 GB RAM over 50 epochs using the Adam optimizer and categorical cross-entropy.
Table \ref{tab6} compares model performance before and after data balancing. Initially, the model achieved strong results on Edge-IIoT (99.11\% accuracy, low loss and FPR) but slightly lower performance on CICIoMT2024 (96.10\% accuracy, higher loss) due to class imbalance. After balancing, both datasets improved; Edge-IIoT reached 99.33\% accuracy with reduced loss and low FPR, while CICIoMT2024 rose to 98.16\% accuracy with notable decreases in loss and FPR. Precision, recall, and F1-scores also increased, and inference time remained low, showing that data balancing enhanced both performance and efficiency.

\noindent \textbf{\textit{{- Accuracy and loss graph:}}} 
The presented results in Figure ~\ref{fig2} with four subfigures (a)-(d) illustrate the training and validation performance of the enhanced SE ViT-BiLSTM model in terms of accuracy and loss. Before balancing: Accuracy (a): The model reaches high accuracy quickly. EdgeIIoT achieves 99.11\%, while CICIoMT2024 is slightly lower at 96.10\%, reflecting class imbalance impact. Loss (b): EdgeIIoT shows low, stable loss (0.0247). CICIoMT2024 shows higher, fluctuating loss (0.1440), indicating less stable learning. After Balancing: Accuracy (c): Accuracy improves and aligns across datasets; 99.33\% for EdgeIIoT and 98.16\% for CICIoMT2024. Loss (d): Loss values decrease and stabilize (0.0158 for EdgeIIoT, 0.0578 for CICIoMT2024), showing more consistent and effective learning.

\begin{figure}[ht]
\centerline{\includegraphics[scale=0.2]{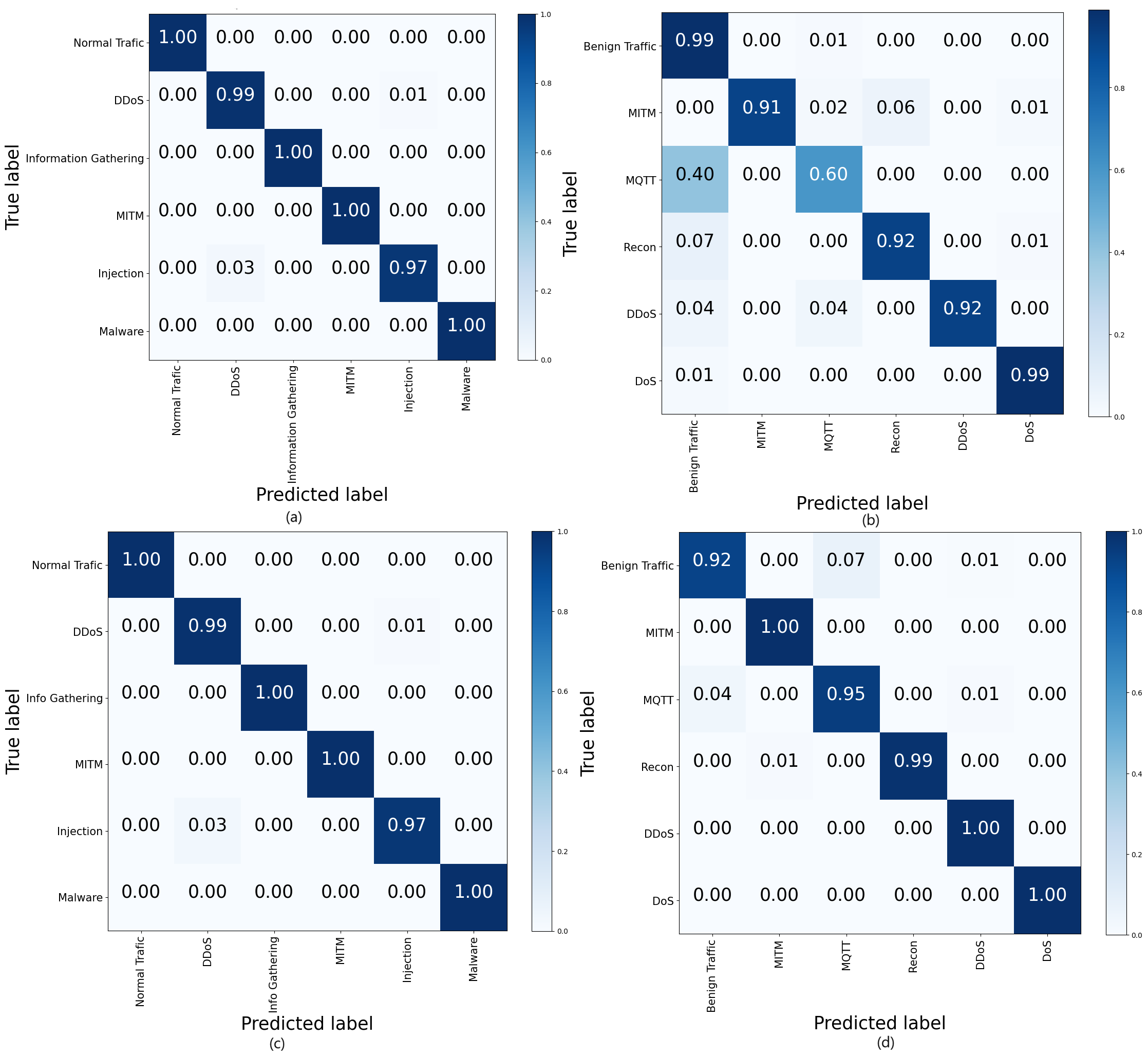}}
\caption{Confusion matrix of the proposed model. (a): Before balancing the EdgeIIoT dataset, (b): Before balancing the CICIoMT2024 dataset, (c): After balancing the EdgeIIoT dataset, (d): After balancing the CICIoMT2024 dataset.}
\label{fig3}
\end{figure}

\noindent \textbf{\textit{{- Confusion matrix:}}} Figure \ref{fig3} (subfigures a-d) presents confusion matrices showing the IDS model’s performance on the EdgeIIoT and CICIoMT2024 datasets before and after data balancing.

Before balancing, the model performs nearly perfectly on EdgeIIoT (subfigure a) but shows noticeable misclassifications on CICIoMT2024 (subfigure b), particularly within the MQTT class. After balancing, EdgeIIoT performance remains consistently high (subfigure c), while CICIoMT2024 shows major improvement (subfigure d), with reduced confusion in the MQTT class and near-perfect classification across all categories, including previously problematic ones like Recon and DDoS. These results highlight the strong positive effect of data balancing on multiclass classification performance.

\begin{figure}[htbp]
\centerline{\includegraphics[scale=0.38]{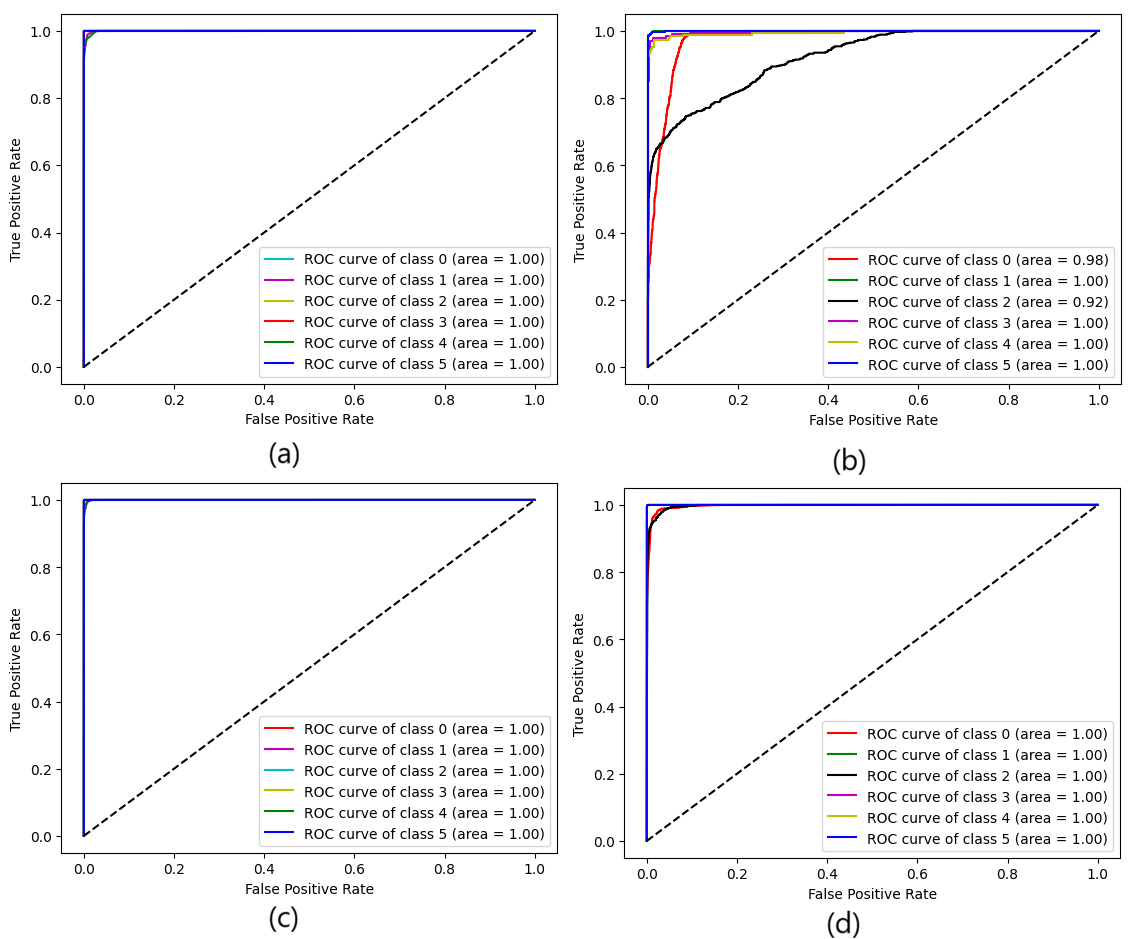}}
\caption{ROC curves of the proposed model. (a): Before balancing EdgeIIoT dataset, (b): Before balancing CICIoMT2024 dataset, (c): After balancing EdgeIIoT dataset, (d): After balancing CICIoMT2024 dataset.}
\label{fig4}
\end{figure}

\noindent \textbf{\textit{{- ROC curve:}}} The presented ROC curves in Figure \ref{fig4}, subfigures (a)-(d) offer a detailed visualization of the classification performance. Before Balancing; on EdgeIIoT, perfect classification with AUC of 1.00 for all classes. However, on CICIoMT2024, most classes perform well (AUC = 1.00), but Class 0 and Class 2 show lower AUCs (0.98 and 0.92), indicating weaker detection, especially for MQTT-related attacks. After Balancing; EdgeIIoT, maintains perfect AUC (1.00) for all classes and CICIoMT2024, Significant improvement; all classes, including the previously weak Class 2, reach AUC = 1.00. Overall, data balancing greatly improves classification fairness and accuracy, particularly for harder-to-detect classes.

\noindent \textbf{\textit{{- Classification report:}}}
Table \ref{cr} summarizes the Pr, Rc, F1 score, and VD per class across two datasets: EdgeIIoTset and CICIoMT2024; before and after data balancing.
Before balancing, the model achieved near-perfect results (99–100\%) across all classes on EdgeIIoTset, but struggled with minority classes in CICIoMT2024. After balancing, EdgeIIoTset metrics remained consistently high, showing the model’s robustness, while CICIoMT2024 exhibited substantial improvements across all metrics, demonstrating better generalization and handling of class imbalance

\begin{table}[ht!]
\centering
\footnotesize
\caption{Classification report for the SE enhanced-ViT-BiLSTM model on EdgeIIoTset and CICIoMT2024 datasets.}
\label{cr}
\small

\begin{tabular}{|m{2.5cm}m{1cm}m{1.cm}m{1.cm}m{1.2cm}|}
% ===== BEFORE DATA BALANCING =====
\hline
\multicolumn{5}{|c|}{\textbf{Before Data Balancing}} \\
\hline

% ================= EdgeIIoT (B.D.B) =================
\multicolumn{5}{|c|}{\textbf{EdgeIIoT Dataset}} \\
& Pr. & Rc & F1& Support \\
\hline
Normal traffic (0) & 100\% & 100\% & 100\% & 4\,985 \\
DDoS (1) & 99\% & 98\% & 99\% & 9\,896 \\
Info. gathering (2) & 100\% & 100\% & 100\% & 4\,333 \\
MITM (3) & 100\% & 100\% & 100\% & 255 \\
Injection (4) & 97\% & 98\% & 98\% & 6\,083 \\
Malware (5) & 100\% & 100\% & 100\% & 6\,008 \\
\hline

% ================= CICIoMT2024 (B.D.B) =================
\multicolumn{5}{|c|}{\textbf{CICIoMT2024 Dataset}} \\
& Pr. & Rc & F1 & Support \\
\hline
Normal traffic & 99\% & 96\% & 94\% & 6\,761 \\
DDoS UDP flood & 91\% & 100\% & 96\% & 472 \\
DoS UDP flood & 60\% & 83\% & 69\% & 455 \\
MITM & 92\% & 86\% & 89\% & 209 \\
MQTT & 92\% & 94\% & 93\% & 181 \\
Recon & 99\% & 99\% & 99\% & 1\,650 \\
\hline
\hline

% ===== AFTER DATA BALANCING =====
\multicolumn{5}{|c|}{\textbf{After Data Balancing}} \\
\hline

% ================= EdgeIIoT (A.D.B) =================
\multicolumn{5}{|c|}{\textbf{EdgeIIoT Dataset}} \\
& Pr. & Rc & F1 & Support \\
\hline
Normal traffic (0) & 100\% & 100\% & 100\% & 9\,860 \\
DDoS (1) & 99\% & 97\% & 98\% & 10\,029 \\
Info. gathering (2) & 100\% & 100\% & 100\% & 9\,878 \\
MITM (3) & 100\% & 100\% & 100\% & 9\,815 \\
Injection (4) & 97\% & 99\% & 98\% & 9\,769 \\
Malware (5) & 100\% & 100\% & 100\% & 9\,925 \\
\hline

% ================= CICIoMT2024 (A.D.B) =================
\multicolumn{5}{|c|}{\textbf{CICIoMT2024 Dataset}} \\
& Pr. & Rc. & F1 & Support \\
\hline
Normal traffic & 92\% & 96\% & 94\% & 6\,284 \\
DDoS UDP flood & 100\% & 99\% & 99\% & 6\,644 \\
DoS UDP flood & 95\% & 93\% & 94\% & 6\,598 \\
MITM & 99\% & 100\% & 99\% & 6\,399 \\
MQTT & 100\% & 98\% & 99\% & 6\,668 \\
Recon & 100\% & 100\% & 100\% & 6\,551 \\
\hline

\end{tabular}

\begin{flushleft}
\vspace{0.1cm}
\footnotesize
Abbreviations: Validation data (Support), Precision (Pr), Recall (Rc).
\end{flushleft}

\end{table}

\begin{table*}[ht]
\centering
\scriptsize
\caption{Performance of different variants of the proposed models in multiclass classification.}
\label{tab70}
\begin{tabular}{|m{0.5cm} | m{2cm} |c| m{0.9cm} |m{0.9cm} |m{0.9cm}||m{0.9cm} |m{0.9cm}|m{0.9cm}|}

\hline
  \multirow{2}{*}{}& \multirow{2}{*}{Model}& \multirow{2}{*} {Description} & \multicolumn{3}{c||}{ EdgeIIoT}  & \multicolumn{3}{c|}{CICIoMT2024}\\
\cline{4-9}
  & & & Acc. (\%)& Loss (\%) & FPR (\%) & Acc. (\%)&Loss (\%)&FPR (\%) \\
\hline

%{\#1} & BiGRU & & & & 99.01&0.0305 &\\ 

\#1 & ViT $\rightarrow$ BiLSTM {$_{32}$} & ViT features passed to a BiLSTM sequentially &98.99& 0.0281 & 0.0023&97.10 & 0.0799& 0.0100\\

\#2 & BiLSTM {$_{32}$} $\rightarrow$ ViT & BiLSTM features passed to a ViT sequentially& 98.93& 0.0286&0.0023 &94.64 & 0.1186&0.0106 \\

\textbf{\#3}  & \textbf{ViT $\parallel$ BiLSTM {$_{32}$}}& \textbf{ViT and BiLSTM {$_{32}$} run in parallel, then fused} & \circled{99.33} & \circled{0.0158}& \circled{0.0013}& \circled{98.16}& \circled{0.0578} &\circled{0.0036}\\

\#4  & ViT $\parallel$ BiLSTM {$_{64}$}& ViT and BiLSTM {$_{64}$} run in parallel, then fused & 98.98 & 0.0301&0.0040 & 96.01& 0.0898 &0.005\\

\hline
\end{tabular}

\end{table*}

\noindent \textbf{\textit{{- Comparative study:}}}  Table \ref{tab5} compares the proposed ViT-BiLSTM model with state-of-the-art approaches for multiclass classification on the EdgeIIoT and CICIoMT2024 datasets. The proposed model achieves the highest accuracy on EdgeIIoT and competitive performance on CICIoMT2024. Although a prior study \cite{kharoubi2025network} reports slightly higher accuracy on CICIoMT2024, it omits key metrics like FPR and loss. Unlike earlier methods that focus mainly on accuracy, the proposed model delivers comprehensive performance; combining high accuracy, full metric reporting, and low inference time.

\begin{table}[ht!]
\centering
\caption{Performance metrics of the proposed enhanced ViT-BiLSTM model in comparison to state-of-the-art methods for multiclass classification.}
%\footnotesize
\scriptsize
%\vspace{0.1cm}
\label{tab5}
\begin{tabular}{|m{0.2cm}m{1.4cm}m{0.7cm}m{0.3cm}m{0.3cm}m{0.2cm}m{0.2cm}m{0.2cm}m{0.3cm}m{0.7cm}|}
\hline
Work  & Model  & Dataset& Acc & Loss & Pr  & Rc & F1 & FPR & Inf time\\
\hline
\hline

\cite{singh2024convolutional} & Generic CNN & EdgeIIoT & 98.98 & \ding{55}  & \ding{55} & \ding{55} & \ding{55} & \ding{55} & \ding{55}\\[0.6mm]

\cite{kavkas2025enhancing} & LSTM/DNN & CICIoMT & 79 & \ding{55}  & 78 & 79 & 76 & \ding{55} & \ding{55}\\[0.6mm]

\hline
\cite{sasi2024efficient}&CNN-LSTM-ResNet-SA& EdgeIIoMT & 33.30 & \ding{55} & 33.31 & 100 & 49.97 & \ding{55} & \ding{55}\\[0.6mm]

& & CICIoT & 99.88 & \ding{55} & 99.89 & 99.99 & 99.94 & \ding{55} & \ding{55}\\[0.6mm]

\cite{kharoubi2025network} & CNN& EdgeIIoT & 96.50 & \ding{55} & 97.48 & 96.50 & 96.41 & \ding{55} & 0.00012 \\[0.6mm]

& & CICIoMT & 99.67 & \ding{55} & 99.67 & 99.67 & 99.66 & \ding{55} & 0.00004\\[0.6mm]
\hline
\hline

\textbf{Our} & \textbf{ViT-BiLSTM} & \textbf{EdgeIIoT} & \textbf{99.33} & \textbf{0.0158} & \textbf{99.33} & \textbf{99.33} & \textbf{99.33} & \textbf{0.0013}& \textbf{0.00035}\\[0.6mm]

\textbf{} & \textbf{} & \textbf{CICIoMT} & \textbf{98.16} & \textbf{0.0578} & \textbf{98.16} & \textbf{98.16} & \textbf{98.15} & \textbf{0.0036}& \textbf{0.00014}\\
\hline
\end{tabular}
\begin{flushleft}
%\vspace{0.0001mm}
 Acc, Pr, Rc, and FPR are in (\%), Inference time is in (seconds/instance).
\end{flushleft} 
\end{table}

\subsection{Ablation Study}

An ablation study (Table \ref{tab70}) was conducted using four model variants to assess the impact of different ViT and BiLSTM configurations in both sequential and parallel setups. Across the EdgeIIoT and CICIoMT2024 datasets, Model~\#3 achieved the best results; 99.33\% accuracy on EdgeIIoT and 98.16\% on CICIoMT2024; along with low loss and FPR. This highlights the effectiveness of parallel feature extraction and fusion, enabling the model to capture both spatial and temporal patterns efficiently. Increasing the BiLSTM size in Model~\#4 did not yield better performance, showing that larger models do not always generalize better. Sequential designs (Models~\#1 and~\#2) performed reasonably well but were outperformed by the parallel architecture.

\section{Conclusion}
\label{sec5}
This study introduced an enhanced SE ViT-BiLSTM-based intrusion detection framework designed to accurately detect a broad range of cyberattacks in both industrial and medical IoT environments. Through rigorous evaluation on two real-world benchmark datasets; EdgeIIoT and CICIoMT2024; the model demonstrated high classification accuracy and real-time performance, both before and after applying class balancing techniques (SMOTE and RandomOverSampler). These results highlight the framework's effectiveness in addressing data imbalance and its suitability for critical IoT applications. Looking ahead, future work will focus on improving the model’s generalizability by testing it on additional datasets, optimizing it for deployment through model compression and edge computing, and enhancing interpretability using explainable AI (XAI) tools to support human decision-making in security operations.

\begin{scriptsize}
\balance

\tiny

%\IEEEtriggeratref{24}
\bibliographystyle{IEEEtran}
\bibliography{references.bib}

\end{scriptsize}

\end{document}